\documentclass[letterpaper,10pt,oneside,conference,final]{IEEEtran}

\usepackage[latin1]{inputenc}
\usepackage{graphicx}
\usepackage{color}
\usepackage{booktabs}
\usepackage{subfig}
\usepackage{tikz,pgfplots}
\usepackage{amssymb}
\usepackage{amsmath}
\usepackage{tabularx}
\usepackage{makecell}
\usetikzlibrary{spy}
\graphicspath{{figures/}}

\begin{document}
\IEEEoverridecommandlockouts
\title{Iterative Receiver for Non-Orthogonal Waveforms Based on the Sum-Product Algorithm}

\author{Ivo Bizon Franco de Almeida$^{1}$, Guilherme Pedro Aquino$^{2}$ and Luciano Leonel Mendes$^{2}$ \\
{\small $^{1}$Vodafone Chair Mobile Communications Systems -- TU Dresden, Dresden, Germany} \\
{\small $^{2}$Instituto Nacional de Telecomunica\c{c}\~oes (Inatel), Santa Rita do Sapuca\'i, Brazil} \\
{\small \texttt{ivo.bizon@ifn.et.tu-dresden.de,}} {\small \texttt{guilhermeaquino@inatel.br,}} {\small \texttt{luciano@inatel.br}}}


\maketitle

\begin{abstract}

~ Based on the application of the Sum-Product algorithm (SPA) over factor graphs, this paper presents a graphical representation of generalized frequency division multiplexing (GFDM) and filter bank multicarrier with offset QAM (FBMC-OQAM).
FBMC-OQAM was chosen because it has the advantage of reducing the algorithm's complexity, since it is directly related to the number of possible values assumed by the transmitted data symbols. 
The receiver algorithm performance is evaluated by the bit error ratio (BER) estimation considering two channel models, additive white Gaussian noise (AWGN) and flat-fading time-variant (Rayleigh).
Likewise, a computational complexity analysis is presented. 
Numerical results show that the BER curves of the proposed scheme present a good match compared with theoretical bit error probability curves.
\end{abstract}


\section{Introduction}

\IEEEPARstart{R}{ecently}, non-orthogonal waveforms, such as filter bank multicarrier with offset quadrature amplitude modulation (FBMC-OQAM) \cite{fbmc_for_next} and generalized frequency division multiplexing (GFDM) \cite{book}, have been considered alternatives for the well employed orthogonal frequency division multiplexing (OFDM) \cite{5gnow}.
These non-orthogonal counterparts present advantages over OFDM.
For instance, FBMC-OQAM presents improved spectral efficiency, low out-of-band emission, and robustness against multiuser interference as advantages.
Similarly, GFDM presents advantages in low latency communications, spectral efficiency, and flexibility with the possibility of adapting its parameters for attending different requirements and covering other waveforms as corner cases \cite{book}.
Non-iterative detection schemes for the mentioned waveforms are well-known \cite{book}\cite{behrouz}.
However, most of these approaches need channel equalization prior to demodulation, e.g., zero forcing, and equalizers often need the noise variance estimation, e.g., minimum mean square error (MMSE) equalizer.

This paper presents an iterative detection algorithm based on the Sum-Product algorithm (SPA) over factor graphs \cite{FG_SP} for estimating complex-valued data symbols transmitted using a modified version of GFDM that emulates FBMC-OQAM. 
This setting is referred as Linear GFDM \cite{Lineargfdm}. 
The algorithm is tailored for Linear GFDM, since it presents benefits in terms of complexity due to the separation of real and imaginary parts of transmitted QAM symbols.
However, the proposed algorithm can be extended for conventional GFDM.

The SPA is an instance of the broader class of  message passing algorithms, and they are often employed for dealing with inference problems that involve estimating marginal probabilities in graphical models.
In fact, if a factor graph is cycle free, the algorithm converges to the exact marginal distribution related to the variables in the graph.
On the other hand, if the graph naturally contains cycles, an iterative schedule shall be used for approaching the exact marginals.

The performance of the proposed receiver algorithm is evaluated through the estimated bit error ratio (BER) under additive white Gaussian noise (AWGN) and flat-fading time-variant (Rayleigh). 
Notably, numerical results and theoretical bit error probability curves present a good match.

The remainder of the paper is organized as follows: 
Section II presents a background on GFDM and FBMC-OQAM.
Section III describes the proposed receiver based on the SPA with some implementation aspects.
Section IV presents the BER performance evaluation of the proposed scheme.
Section V concludes the paper.

\section{Non-orthogonal Waveforms}

GFDM and FBMC are non-orthogonal multicarrier waveforms, and, hence, present intrinsic intersymbol (ISI) and/or intercarrier (ICI) interference. 
However, according to the Balian-Low Theorem \cite{balianlow}, setting aside full orthogonality leads to a new degree of freedom when projecting the waveform characteristics, such as time-frequency localization. 
The following subsections present a short background on FBMC-OQAM and GFDM and briefly demonstrates how to emulate FBMC-OQAM using the GFDM matrix formulation.
	
\subsection{FBMC-OQAM}

The Balian-Low Theorem states that it is impossible to synthesize a waveform that presents at the same time the following characteristics: \textit{i)} orthogonality in the complex field, \textit{ii)} good time and frequency localization and \textit{iii)} operates at the Nyquist rate, i.e., $R = 1/T$, where $R$ represents the data symbol rate and $T$ its time spacing.
OFDM lacks in time-frequency localization, which often leads to undesired levels of out-of-band emission \cite{5gnow}.
FBMC-OQAM presents \textit{ii)} and \textit{iii)}. 
However, only it presents real orthogonality, i.e., the real part of a symbol suffers from interference from the imaginary part and vice-versa.
For overcoming this limitation OQAM is often employed in conjunction with FBMC \cite{fbmc_for_next}.
OQAM avoids ICI by introducing a $\pi/2$ phase rotation among adjacent subcarriers, and a time shift between imaginary and real parts from the transmitted symbol.

Ideally, FBMC-OQAM continuously sends data symbols over $K$ subcarriers.
Thus, the discrete-time transmit signal is described as
\begin{equation}
x[n] = \sum_{m=-\infty}^{+\infty} \sum_{k=0}^{K-1} d_{k,m}^{\mathcal{(I)}} g^{\mathcal{(I)}}_{k,m}\left[n\right] + j d_{k,m}^{\mathcal{(Q)}} g^{\mathcal{(Q)}}_{k,m}[n],
\end{equation}
where $d_{k,m} = d_{k,m}^{\mathcal{(I)}} + jd_{k,m}^{\mathcal{(Q)}}$ represents the QAM data symbol sent through the $k$th subcarrier in the $m$th time slot, and $g^{\mathcal{(I)}}_{k,m}[n]$ and $g^{\mathcal{(Q)}}_{k,m}[n]$ are respectively given by 
\begin{equation}
g^{\mathcal{(I)}}_{k,m}[n] = p\left[ n-mK\right] \exp\left( {j2\pi \frac{k}{K}n + j\frac{\pi}{2}k}\right) 
\end{equation}
\begin{equation}
g^{\mathcal{(Q)}}_{k,m}[n] = p\left[ n-mK-\frac{K}{2}\right] \exp \left( {j2\pi \frac{k}{K}n + j\frac{\pi}{2}k}\right)  ,
\end{equation}
where $p[n]$ represents the prototype filter impulse response.

\subsection{GFDM}

Similarly to FBMC, GFDM is also based on a prototype filter.
However, it employs circular filtering for shaping data symbols, which are transmitted in $K$ subcarriers and $M$ time slots, referred as subsymbols.
Hence, one GFDM frame carriers $N=KM$ QAM data symbols \cite{book}.

Matrix formulation can be used for describing the transmit signal.
Thus, the transmit vector is given by
\begin{equation}
\mathbf{x=Ad},
\end{equation} 
where $\mathbf{d}$ represents the $N\times1$ data symbol vector, and $\mathbf{A}$ represents the  $N\times N$ transmit matrix.
The transmit matrix is assembled as
\begin{equation} \label{matA}
\mathbf{A}=[\mathbf{g}_{0,0} \ \mathbf{g}_{1,0} \cdots \mathbf{g}_{K-1,0} \; \cdots \; \mathbf{g}_{0,M-1} \cdots \mathbf{g}_{K-1,M-1}]
\end{equation}
where $\mathbf{g}_{k,m}$ represents the vector with samples from the prototype filter modulated on the $k$th subcarrier and circularly shifted to the $m$th subsymbol.

The received vector can be expressed as
\begin{equation}\label{y}
	\mathbf{y=\Psi d+w},
\end{equation}
where  $\mathbf{w}$ is the additive white Gaussian noise (AWGN) vector with zero mean and variance $\sigma^{2}$, and the \textit{equivalent} matrix is  given by
\begin{equation}
	\mathbf{\Psi} :=  \mathbf{HA},
\end{equation}
where $\mathbf{H}$ represents the linear Toeplitz matrix from the channel impulse response.

Notably, GFDM is a frame generator for other waveforms.
For instance, if $M=1$ and rectangular filter is chosen as the prototype filter, the resulting waveform is OFDM.
Analogously, the next subsection explores this flexibility for generating FBMC-OQAM from GFDM.

\subsection{Linear GFDM}

GFDM displays circular filtering behavior, whereas FBMC displays linear behavior.
For achieving such linear filtering in GFDM it is necessary to zero pad the prototype filter \cite{Lineargfdm}.
The length of the zeroed sequence that is padded to the prototype filter impulse response is given by 
\begin{equation}
L_Z = KM - K/2,
\end{equation} 
where $M$ is total number of subsymbols, which can be translated as the overlapping factor in the FBMC-OQAM context.

OQAM can be created by employing two modulation matrices $\mathbf{A^{\mathrm{(L)}}_{\mathrm{I}}}$ and $\mathbf{A^{\mathrm{(L)}}_{\mathrm{Q}}}$, where one is $K/2$ samples shifted in relation to the other. 
Both matrices are assembled similarly to (\ref{matA}), with the samples from the zero padded prototype filter.

The transmit vector is obtained by adding the in-phase component with the quadrature component, yielding to
\begin{equation}
\begin{split}
	\mathbf{x} & = \mathbf{A^{\mathrm{(L)}}_{\mathrm{I}}} \Re \left\lbrace \mathbf{d} \right\rbrace  + j \mathbf{A^{\mathrm{(L)}}_{\mathrm{Q}}} \Im \left\lbrace \mathbf{d} \right\rbrace\\
	           & = \mathbf{x}_{\mathrm{I}} + \mathbf{x}_{\mathrm{Q}},
\end{split}
\end{equation}

In this case, the received vector can be expressed as
\begin{equation}
\begin{split}
	\mathbf{y} & = \mathbf{\Psi_{\mathrm{I}} \Re \left\lbrace \mathbf{d} \right\rbrace + \mathit{j}\Psi_{\mathrm{Q}} \Im \left\lbrace \mathbf{d} \right\rbrace + w} \\
	           & = \mathbf{y}_{\mathrm{I}} + \mathbf{y}_{\mathrm{Q}} + \mathbf{w},
\end{split}
\end{equation}

Since intrinsic ICI is eliminated by OQAM, half-Nyquist pulses can be employed as prototype filter for eliminating ISI with matched filtering demodulation.
However, dispersive channels ruin the orthogonality created by OQAM, and equalization prior to demodulation becomes necessary.
The next section describes a non-linear algorithm that demodulates data symbols without the need of prior equalization and knowledge of the signal-to-noise ratio.
	
\section{Sum-Product Algorithm Based Receiver}
\begin{figure*}[ht!]
	\centering
	\begin{tikzpicture}[scale=0.9, every node/.style={scale=0.9}]
	\draw (4,-1.2) circle (0.5);
	\draw (6,-1.2) circle (0.5);
	\draw (8,-1.2) circle (0.5);	
	\draw (10,-1.2) circle (0.5);
	\draw (12,-1.2) circle (0.5);
	\draw (14,-1.2) circle (0.5);	
	
	\node at (4,-1.2) {$\mathbf{d}_{\mathrm{I}}^{(1)}$};
	\node at (6,-1.2) {$\mathbf{d}_{\mathrm{I}}^{(2)}$};
	\node at (8,-1.2) {$\mathbf{d}_{\mathrm{I}}^{(3)}$};			
	\node at (10,-1.2) {$\mathbf{d}_{\mathrm{I}}^{(4)}$};
	\node at (12,-1.2) {$\mathbf{d}_{\mathrm{I}}^{(5)}$};
	\node at (14,-1.2) {$\mathbf{d}_{\mathrm{I}}^{(6)}$};	
	
	\draw [fill=black] (1.75,1.75) rectangle (2.25,2.25);
	\draw [fill=black] (3.25,1.75) rectangle (3.75,2.25);
	\draw [fill=black] (4.75,1.75) rectangle (5.25,2.25);	
	\draw [fill=black] (6.25,1.75) rectangle (6.75,2.25);
	\draw [fill=black] (7.75,1.75) rectangle (8.25,2.25);	
	\draw [fill=black] (9.25,1.75) rectangle (9.75,2.25);	
	\draw [fill=black] (10.75,1.75) rectangle (11.25,2.25);	
	\draw [fill=black] (12.25,1.75) rectangle (12.75,2.25);	
	\draw [fill=black] (13.75,1.75) rectangle (14.25,2.25);						
	\draw [fill=black] (15.25,1.75) rectangle (15.75,2.25);	
	\draw [fill=black] (16.75,1.75) rectangle (17.25,2.25);	
	
	\node at (1.5,2) {$f_1$};
	\node at (3,2) {$f_2$};
	\node at (4.5,2) {$f_3$};
	\node at (6,2) {$f_4$};
	\node at (7.5,2) {$f_5$};
	\node at (9,2) {$f_6$};
	\node at (10.5,2) {$f_7$};
	\node at (12,2) {$f_8$};
	\node at (13.45,2) {$f_{9}$};
	\node at (14.95,2) {$f_{10}$};
	\node at (16.45,2) {$f_{11}$};
	
	\node at (2,3) {$\mathbf{y}_{\mathrm{I}}^{(1)}$};
	\node at (3.5,3) {$\mathbf{y}_{\mathrm{I}}^{(2)}$};	
	\node at (5,3) {$\mathbf{y}_{\mathrm{I}}^{(3)}$};	
	\node at (6.5,3) {$\mathbf{y}_{\mathrm{I}}^{(4)}$};	
	\node at (8,3) {$\mathbf{y}_{\mathrm{I}}^{(5)}$};	
	\node at (9.5,3) {$\mathbf{y}_{\mathrm{I}}^{(6)}$};	
	\node at (11,3) {$\mathbf{y}_{\mathrm{I}}^{(7)}$};	
	\node at (12.5,3) {$\mathbf{y}_{\mathrm{I}}^{(8)}$};	
	\node at (14,3) {$\mathbf{y}_{\mathrm{I}}^{(9)}$};	
	\node at (15.5,3) {$\mathbf{y}_{\mathrm{I}}^{(10)}$};	
	\node at (17,3) {$\mathbf{y}_{\mathrm{I}}^{(11)}$};
	
	\draw (2,3) circle (0.5);
	\draw (3.5,3) circle (0.5);
	\draw (5,3) circle (0.5);
	\draw (6.5,3) circle (0.5);
	\draw (8,3) circle (0.5);
	\draw (9.5,3) circle (0.5);
	\draw (11,3) circle (0.5);
	\draw (12.5,3) circle (0.5);
	\draw (14,3) circle (0.5);
	\draw (15.5,3) circle (0.5);
	\draw(17,3) circle (0.5);
	
	\draw (2,2.25)--(2,2.5);
	\draw (3.5,2.25)--(3.5,2.5);
	\draw (5,2.25)--(5,2.5);
	\draw (6.5,2.25)--(6.5,2.5);
	\draw (8,2.25)--(8,2.5);
	\draw (9.5,2.25)--(9.5,2.5);
	\draw (11,2.25)--(11,2.5);
	\draw (12.5,2.25)--(12.5,2.5);
	\draw (14,2.25)--(14,2.5);
	\draw (15.5,2.25)--(15.5,2.5);
	\draw (17,2.25)--(17,2.5);
	
	\draw (5,1.75)--(12,-0.7);	
	\draw (5,1.75)--(14,-0.7);
	
	\draw (6.5,1.75)--(12,-0.7);	
	\draw (6.5,1.75)--(14,-0.7);
	
	\draw (8,1.75)--(8,-0.7);	
	\draw (8,1.75)--(10,-0.7);
	\draw (8,1.75)--(12,-0.7);	
	\draw (8,1.75)--(14,-0.7);	
	
	\draw (9.5,1.75)--(8,-0.7);
	\draw (9.5,1.75)--(10,-0.7);
	\draw (9.5,1.75)--(12,-0.7);
	\draw (9.5,1.75)--(14,-0.7);				
	
	\draw (11,1.75)--(4,-0.7);
	\draw (11,1.75)--(6,-0.7);
	\draw (11,1.75)--(8,-0.7);
	\draw (11,1.75)--(10,-0.7);
	\draw (11,1.75)--(12,-0.7);
	\draw (11,1.75)--(14,-0.7);	
	
	\draw (12.5,1.75)--(4,-0.7);	
	\draw (12.5,1.75)--(6,-0.7);	
	\draw (12.5,1.75)--(8,-0.7);	
	\draw (12.5,1.75)--(10,-0.7);	
	
	\draw (14,1.75)--(4,-0.7);	
	\draw (14,1.75)--(6,-0.7);	
	\draw (14,1.75)--(8,-0.7);	
	\draw (14,1.75)--(10,-0.7);	
	
	\draw (15.5,1.75)--(4,-0.7);		
	\draw (15.5,1.75)--(6,-0.7);			
	
	\draw (17,1.75)--(4,-0.7);		
	\draw (17,1.75)--(6,-0.7);
	\end{tikzpicture}
	\caption{Factor graph showing the relation among $\mathbf{y_{\mathrm{I}}}$ and $\mathbf{d}_{\mathrm{I}} = \Re \left\lbrace \mathbf{d} \right\rbrace $ over single tap channel.}
	\label{fg}
\end{figure*}
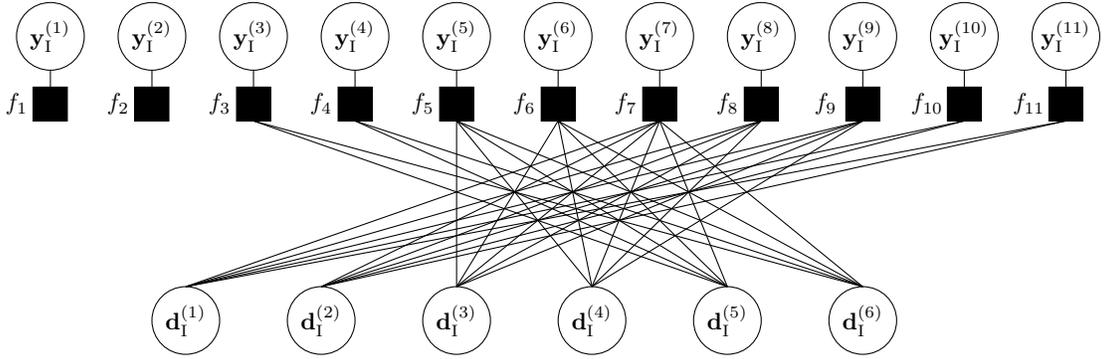

The SPA comes from the class of message passing algorithms that operate over factor graphs.
Factor graphs can be defined as a graphical representation of the relation among a set of variables in a probabilistic model.
It is a bipartite graph composed by variable nodes and function (or factor) nodes.
Variable nodes are represented by circles and factor nodes are represented by filled squares.
Many well known algorithms in coding and estimation theory may be viewed as specific instances of message passing on factor graphs \cite{FG_SP}.

For Linear GFDM, there is one particular graph for each equivalent matrix $\mathbf{\Psi_{\mathrm{I}}}$ and $\mathbf{\Psi_{\mathrm{Q}}}$.
The non-zero values of each matrix determine the edges that connect factor nodes to variable nodes.
Consequently, real and imaginary parts are estimated separately.
The SPA will estimate marginal probabilities related to each variable node of the graph.
Hence, it is possible to use the maximum a posteriori criterion (MAP) for estimating the most probable transmitted data symbol.

The factor graph that represents the relation among the real part of the transmit data symbol vector and in-phase component of the received vector is illustrated in Fig. \ref{fg} considering $K=2$ subcarriers, $M=3$ subsymbols.
The graph for the imaginary part and quadrature component is constructed similarly.
For that configuration the resultant transmit vector has length given by $L_{\mathbf{x}} = KM+KM-K/2$, i.e., the length of the data vector plus the zeroed sequence.
Hence, for the given example the graph contains eleven factor nodes from the received signal samples and six variable nodes from the data vector.

For this paper, full channel state information (CSI) is assumed available at the receiver. 
Messages from the factor node $f_{i}$ to variable node $\mathbf{d}^{(j)}$ over the $t$th iteration are calculated as follows

\begin{IEEEeqnarray}{rCl}
	\IEEEeqnarraymulticol{3}{l}{
		\mu^{(t)}_{f_i\rightarrow \mathbf{d}^{(j)}}\left(\mathbf{d}^{(j)}\right) =
	}\nonumber\\\quad
	\sum_{\sim \left\lbrace \mathbf{d}^{(j)}\right\rbrace } \left\lbrace \exp\left( - \bigg\| \mathbf{y}^{(i)} - \sum_{k \in N(f_i)} \mathbf{\Psi}^{(i,k)}\mathbf{d}^{(k)} \bigg\| ^{2} \right) \right\rbrace \nonumber\\
		\times \prod_{u \in N(f_i)\backslash j}^{} \mu^{(t-1)}_{\mathbf{d}^{(u)} \rightarrow f_i} \left( \mathbf{d}^{(u)} \right),
	\nonumber\\
\end{IEEEeqnarray}
where $i$th sample from the received vector is represented by $\mathbf{y}^{(i)}$, the $i$th factor node is represented by $f_i$, and $\mathbf{\Psi}^{(i,k)}$ represents the element in the $i$th line and $k$th column of $\mathbf{\Psi}$.
The notation $N(f_i)\backslash j$ represents the set of variable nodes connected to $f_i$ excluding the $j$th node. 
Messages must be calculated for the in-phase and quadrature components.
However, in (11) the subindex $(\mathrm{I})$ and $(\mathrm{Q})$ were removed for the sake of brevity.
The notation $\sum_{\sim \left\lbrace \cdot \right\rbrace }^{}$ represents the \textit{summary} operation described in \cite{FG_SP}.
The sum in (11) is carried over an alphabet of possible values assumed by $\mathbf{d}^{(k)}$.
The alphabets for the in-phase and quadrature components are described as 
\begin{gather}
\mathbb{D}_{\mathrm{I}}  \in \Re \left\lbrace \mathbb{D} \right\rbrace \\
\mathbb{D}_{\mathrm{Q}}  \in \Im \left\lbrace \mathbb{D} \right\rbrace 	
\end{gather}
where $\mathbb{D}$ is the the $J$-QAM mapping alphabet.

Messages from variable node  $\mathbf{d}^{(j)}$ to factor node $f_i$ over the $t$th iteration are given by
\begin{equation}\label{v2f}
\mu^{(t)}_{\mathbf{d}^{(j)} \rightarrow f_i}\left( \mathbf{d}^{(j)}\right)  = \prod_{u \in N(\mathbf{d}^{(j)}) \backslash i}^{} \mu^{(t-1)}_{f_u \rightarrow \mathbf{d}^{(j)}} \left( \mathbf{d}^{(j)}\right),
\end{equation}
where $N(\mathbf{d}^{(j)}) \backslash i$ represents the set of factor nodes connected to $\mathbf{d}^{(j)}$ excluding the $i$th node.

After $\tau$ iterations, the non-normalized marginal probability distribution of each data component is given by the multiplication of all incoming messages at the variable nodes $\mathbf{d}^{(j)}$, as follows
\begin{equation}\label{marginal}
	p \left( \mathbf{d}^{(j)}\right)  = \prod_{u\in N(\mathbf{d})}^{} \mu^{(\tau)}_{f_u \rightarrow \mathbf{d}^{(j)}} \left( \mathbf{d}^{(j)}\right).
\end{equation}
Thus, $p\left( \mathbf{d}^{(j)}\right)$ is the probability mass function that contains the likelihood of each possible value assumed by $\mathbf{d}^{(j)}$, and MAP criterion can be used for estimating the received data components.

It is important to emphasize that messages must be calculated for the in-phase and quadrature components, leading to two different algorithms that can operate in parallel.
This characteristic leads to a less complex algorithm since the message computation complexity is directly related to transmit data symbols alphabet.
In fact, the number of possible values is reduced by a factor of 2 since the algorithm is dealing separately with real and imaginary parts of QAM symbols.
Hence, with this separation complexity is reduced when compared with an algorithm designed to estimate complex-valued symbols.
Another key point of this demodulating approach lies in the fact that estimation of noise variance and equalization prior to demodulation are not necessary for message computation.

Considering the graph shown in Fig. \ref{fg}, one can see that it is a cyclic graph with girth equals four, and iterative message passing becomes necessary for convergence.
Although the resulting marginal probabilities will not be exact, numerical results show that it can deliver acceptable BER performance under AWGN and Rayleigh channels.

\section{Performance Evaluation}

Firstly, for evaluating the performance of the proposed receiver algorithm we resort to the BER under two different channel models was estimated through Monte Carlo simulation.
The simulation parameters are shown in Table \ref{tabela1}.
Complexity analysis in terms of number of complex multiplications is also presented in this section. 	
\begin{table}[h]
	\centering
	\caption{Waveform simulation parameters.}
	\label{tabela1}
	\renewcommand{\arraystretch}{1.3}
	\begin{tabularx}{7.5cm}{XX}
		\toprule[0.9pt] 
		Parameter                & Value                         \\ 
		\midrule 
		Waveform				 & Linear GFDM                   \\
		Mapping                  & QPSK                          \\ 
		Prototype filter         & Martin \cite{martin_filter}   \\ 
		Number of subcarriers    & $K=2$                         \\ 
		Number of subsymbols     & $M=3$                         \\
		Number of iterations     & $\tau = 1 $ and $  7$         \\
		\bottomrule[0.9pt]	
	\end{tabularx}
\end{table}

\subsection{BER}
\begin{figure}[t!]
	\centering
	\begin{tikzpicture}[scale=1, every node/.style={scale=1}, spy using outlines={circle, magnification=3.25, size=2.5cm}, connect spies, font=\small]
	\begin{axis}[%
	width=8.82cm,
	height=8.6cm,
	at={(1.453in,1.076in)},
	separate axis lines,
	every outer x axis line/.append style={black},
	every x tick label/.append style={font=\small},
	every x tick/.append style={black},
	xmin=0,
	xmax=40,
	xtick={0,5,10,15,20,25,30,35,40},
	xlabel={$E_{b}/N_{0} \mathrm{(dB)}$},
	every outer y axis line/.append style={black},
	every y tick label/.append style={font=\small},
	every y tick/.append style={black},
	ymode=log,
	ymin=1e-006,
	ymax=1,
	yminorticks=true,
	ytick={1,1e-1,1e-2,1e-3,1e-4,1e-5,1e-6,1e-7},
	ylabel={$\mathrm{BER}$},
	axis background/.style={fill=white},
	xmajorgrids,
	ymajorgrids,
	yminorgrids,
	ylabel near ticks,
	legend style={at={(0.592,0.8)}, anchor=south west, legend cell align=left, align=left, draw=black}
	]
	\addplot [color=blue, line width=0.7pt]
	table[row sep=crcr]{%
		0	0.128957501705917\\
		0.5	0.120450492234267\\
		1	0.112173582183927\\
		1.5	0.104163982837102\\
		2	0.096454225274435\\
		2.5	0.0890716425084265\\
		3	0.0820380574568693\\
		3.5	0.0753696765317705\\
		4	0.0690771758141298\\
		4.5	0.063165956391648\\
		5	0.0576365381552048\\
		5.5	0.05248505741687\\
		6	0.0477038329450743\\
		6.5	0.0432819669136564\\
		7	0.0392059511471821\\
		7.5	0.0354602541730498\\
		8	0.0320278702652264\\
		8.5	0.0288908173051516\\
		9	0.0260305754640743\\
		9.5	0.0234284631560231\\
		10	0.0210659502900651\\
		10.5	0.018924911544509\\
		11	0.0169878242535627\\
		11.5	0.0152379166464695\\
		12	0.0136592727410198\\
		12.5	0.0122369003018566\\
		13	0.0109567680542369\\
		13.5	0.00980581790437796\\
		14	0.00877195734694469\\
		14.5	0.00784403660750807\\
		15	0.00701181442353814\\
		15.5	0.00626591574672031\\
		16	0.00559778407443938\\
		16.5	0.00499963060174581\\
		17	0.00446438193234316\\
		17.5	0.0039856276984245\\
		18	0.00355756911150681\\
		18.5	0.00317496919473996\\
		19	0.00283310522544289\\
		19.5	0.00252772373844015\\
		20	0.00225499829984764\\
		20.5	0.00201149015138564\\
		21	0.00179411174174707\\
		21.5	0.00160009309928905\\
		22	0.00142695095527745\\
		22.5	0.00127246049561333\\
		23	0.00113462959849672\\
		23.5	0.001011675403427\\
		24	0.000902003051315478\\
		24.5	0.000804186434701301\\
		25	0.000716950799826019\\
		25.5	0.000639157047618827\\
		26	0.000569787587671802\\
		26.5	0.000507933607419631\\
		27	0.000452783627508016\\
		27.5	0.000403613223374399\\
		28	0.000359775802113285\\
		28.5	0.000320694332555674\\
		29	0.000285853935030098\\
		29.5	0.000254795245392056\\
		30	0.000227108475562149\\
		30.5	0.000202428099958421\\
		31	0.000180428103846345\\
		31.5	0.000160817735754118\\
		32	0.000143337711727975\\
		32.5	0.000127756824354383\\
		33	0.000113868914171381\\
		33.5	0.000101490165367327\\
		34	9.04566915390231e-05\\
		34.5	8.06223807922057e-05\\
		35	7.18569726385065e-05\\
		35.5	6.40443420038689e-05\\
		36	5.70809682400031e-05\\
		36.5	5.08745693509893e-05\\
		37	4.53428837306619e-05\\
		37.5	4.0412583576635e-05\\
		38	3.60183058284724e-05\\
		38.5	3.21017879806873e-05\\
		39	2.86110974694748e-05\\
		39.5	2.54999445418874e-05\\
		40	2.27270695935811e-05\\
	};
	\addlegendentry{$\mathrm{Theo.\; OFDM}$}
	
	\addplot [color=blue, line width=0.7pt, forget plot]
	table[row sep=crcr]{%
		0	0.0755567234578115\\
		0.5	0.0648163279161176\\
		1	0.0546981229173966\\
		1.5	0.0453247367508956\\
		2	0.0368027735266879\\
		2.5	0.0292155695839453\\
		3	0.0226166967948198\\
		3.5	0.0170250935853833\\
		4	0.0124226828148937\\
		4.5	0.00875514497873712\\
		5	0.00593614288077196\\
		5.5	0.00385477322617877\\
		6	0.00238543881450566\\
		6.5	0.00139882511268592\\
		7	0.000772376302193284\\
		7.5	0.000398716815900694\\
		8	0.000190889551186892\\
		8.5	8.39960112177196e-05\\
		9	3.36266630243719e-05\\
		9.5	1.21088199643855e-05\\
		10	3.87210071891102e-06\\
		10.5	1.08384783437937e-06\\
		11	2.61306761216899e-07\\
		11.5	5.32866389433742e-08\\
		12	9.00601031007464e-09\\
		12.5	1.23302844585063e-09\\
		13	1.33293101744122e-10\\
		13.5	1.10545978569217e-11\\
		14	6.81018912877844e-13\\
		14.5	3.00554978904591e-14\\
		15	9.12395736262809e-16\\
		15.5	1.82023424951961e-17\\
		16	2.26739584445441e-19\\
		16.5	1.66510221001549e-21\\
		17	6.75896977065469e-24\\
		17.5	1.41072520944947e-26\\
		18	1.39601431090675e-29\\
		18.5	5.9797823640767e-33\\
		19	1.00107397357086e-36\\
		19.5	5.84057298723834e-41\\
		20	1.04424379188127e-45\\
		20.5	4.95265997798362e-51\\
		21	5.29969722887031e-57\\
		21.5	1.06695239928015e-63\\
		22	3.29608811928451e-71\\
		22.5	1.24305092158265e-79\\
		23	4.42763849843474e-89\\
		23.5	1.11688064254648e-99\\
		24	1.44441906577261e-111\\
		24.5	6.66518938940689e-125\\
		25	7.30696918464799e-140\\
		25.5	1.20583575043362e-156\\
		26	1.79516322712013e-175\\
		26.5	1.35734754235084e-196\\
		27	2.73597996566879e-220\\
		27.5	7.13305048496119e-247\\
		28	1.06846073755641e-276\\
		28.5	3.69948468992818e-310\\
		29	0\\
		29.5	0\\
		30	0\\
		30.5	0\\
		31	0\\
		31.5	0\\
		32	0\\
		32.5	0\\
		33	0\\
		33.5	0\\
		34	0\\
		34.5	0\\
		35	0\\
		35.5	0\\
		36	0\\
		36.5	0\\
		37	0\\
		37.5	0\\
		38	0\\
		38.5	0\\
		39	0\\
		39.5	0\\
		40	0\\
	};
	
	\addplot [color=black, densely dotted, line width=0.7pt, mark size=1.7pt, mark=square, mark options={solid, black}]
	table[row sep=crcr]{%
		0	0.132337387625595\\
		2	0.098568426582611\\
		4	0.0713980028530671\\
		6	0.0477531665711922\\
		8	0.0334067547723935\\
		10	0.0214092140921409\\
		12	0.0137650447673565\\
		14	0.00888269280860032\\
		16	0.00567635865658881\\
		18	0.00345868023861551\\
		20	0.00229096647954309\\
		22	0.00137539919116811\\
		24	0.00097270717172267\\
		26	0.000597788957207627\\
		28	0.000346164337861923\\
		30	0.000227113731836577\\
		32	0.000143973606309593\\
		34	8.76540291966117e-05\\
		36	5.60312115484878e-05\\
		38	3.65847350045724e-05\\
		40	2.193543897857445e-05\\
	};
	\addlegendentry{$\tau = 7$}
	
	\addplot [color=black, line width=0.7pt, dotted, mark size=1.7pt, mark=square, mark options={solid, black}, forget plot]
	table[row sep=crcr]{%
		0	0.0760407171072622\\
		1	0.0550605060506051\\
		2	0.0373898102495144\\
		3	0.0242631374830328\\
		4	0.0137787688580553\\
		5	0.00669798190675017\\
		6	0.00305824412181649\\
		7	0.00109618863643332\\
		8	0.00029621975066642\\
		9	6.55592974348697e-05\\
		10	1.11086947942814e-05\\
		11	1.35910827229058e-06\\
		12	0\\
		13	0\\
		14	0\\
		15	0\\
	};
	
	\addplot [color=red, line width=0.7pt, dashed, mark size=1.7pt, mark=o, mark options={solid, red}]
	table[row sep=crcr]{%
		0	0.0767402637227844\\
		1	0.0547712847450208\\
		2	0.0406712172923777\\
		3	0.0264395139989435\\
		4	0.0132596831452339\\
		5	0.0069886617515639\\
		6	0.00307799220200977\\
		7	0.00116492335479306\\
		8	0.000356190744591665\\
		9	8.7670159810177e-05\\
		10	1.50063929332692e-05\\
		11	1.80664842143074e-06\\
		12	1.3601754977706e-07\\
		13	0\\
		14	0\\
		15	0\\
	};
	\addlegendentry{$\tau = 1$}
	
	\addplot [color=red, dashed, line width=0.7pt, mark size=1.7pt, mark=o, mark options={solid, red}, forget plot]
	table[row sep=crcr]{%
		0	0.130635335073977\\
		2	0.096001279590531\\
		4	0.0717632434499904\\
		6	0.0488730375871279\\
		8	0.0331456953642384\\
		10	0.0215491259795057\\
		12	0.0139066525792878\\
		14	0.0088596025709063\\
		16	0.00562901001379145\\
		18	0.0034758875493083\\
		20	0.00228780285571448\\
		22	0.00137849926737102\\
		24	0.000973507364399929\\
		26	0.000597920360114113\\
		28	0.000344925685864389\\
		30	0.000227540653272931\\
		32	0.000143965085324618\\
		34	8.76540291966117e-05\\
		36	5.60125593475729e-05\\
		38	3.68810358131694e-05\\
		40	2.31965150025663e-05\\
	};	
	\end{axis}
	\spy[black] on (5.495,4) in node at (7.3,4.4);
	\spy[black] on (10.4,4.65) in node at (9.5,7);
	\node at (4.6,4.2) {$\mathrm{AWGN}$};
	\draw [->] (4.6,4.5)--(5,5.1);
	\node at (9.3,4.2) {$\mathrm{Rayleigh}$};	
	\draw [->] (9.3,4.5)--(9.4,5);
	
	\end{tikzpicture}%
	\caption{BER performance of the algorithm considering $\tau = 1$ iteration and $\tau = 7$ iterations.}
	\label{ber}
\end{figure}
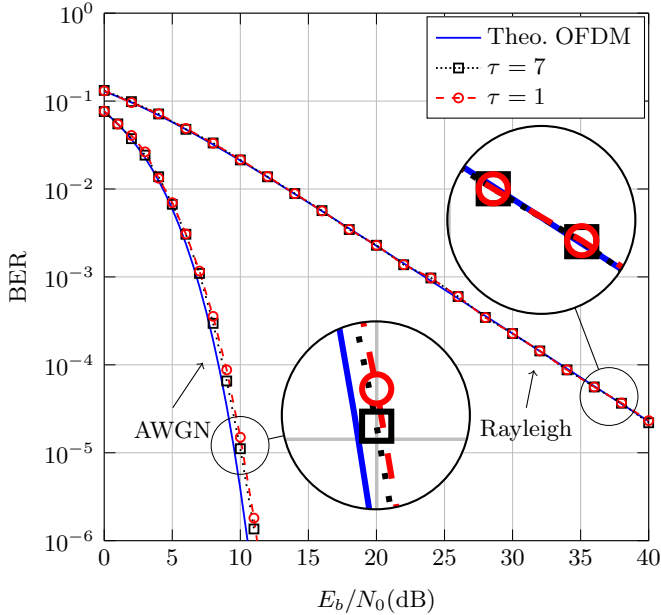

Figure \ref{ber} shows the estimated BER of the proposed scheme. 
For the Rayleigh channel, CSI is assumed to be available at the receiver. 
Although the girth of the graph equals four, one can observe that the proposed receiver algorithm holds acceptable performance, which is the same as theoretical OFDM bit error probability under the simulation assumptions.
This phenomenon is attributed to the random nature of the channel, since the edge values vary with the channel gain at each frame transmission. 
Therefore, the performance loss expected from the cycles in the graph is not present in this scenario.
Moreover, note that the short girth does not degrade performance even for just a single iteration.

For the AWGN channel, edge values are fixed, and due to the short girth, performance degradation is observed.
For $\tau = 7$ iterations, performance under AWGN channel was slightly improved. 

\subsection{Complexity}

For the complexity analysis, we take into account the number of complex multiplications performed for estimating one $J$-QAM data symbol vector whose length is given by $N = KM$.
For computing the messages described by (11), one needs $N-1$ \textit{for} loops.
These loops' index run from $1$ to $J/2$ because real and imaginary parts of the QAM symbol are estimated independently.
Taking into consideration that the algorithm performs message calculations iteratively, and adding the computation of (\ref{v2f}).
The algorithm built in such manner that the number of complex multiplications required is given by
\begin{multline}
	\mathcal{C} = \bigg[ \big( 8N^3 - 4N^2K - 8N^2 + 2NK + 2N \big)\big(J/2\big)^{N-1} \bigg. \\ \bigg. + \ 2N^3 - 8N^2 + 6N \bigg] \tau.
\end{multline}
Therefore, the complexity is $\mathcal{O}(c^N)$.

\section{Conclusion}

In this paper an iterative demodulation algorithm for Linear GFDM was described employing the SPA over the factor graph representation.
The graph that represents the relation among data symbols and received waveform samples is also derived.
Numerical results show that the proposed algorithm holds the same performance as OFDM without the need of previous channel equalization and noise variance estimation when transmission over Rayleigh fading channel is assumed.
Hence, it presents an alternative to the well known MMSE estimator.
Nevertheless, further investigation for designing less complex message computation algorithms that avail redundant calculation remains an open issue.

\section*{Acknowledgments}
This work was supported by RNP, with resources from MCTIC, Grant No. 01250.075413/2018- 04, under the Centro de Refer\^encia em Radiocomunica\c{c}\~oes (CRR) project of the Instituto Nacional de Telecomunica\c{c}\~oes (Inatel), Brazil; with resources from 5G-RANGE BR-EU project; and by CNPq-Brazil.

\bibliographystyle{IEEEtran}
\bibliography{IEEEabrv,my_references}
\end{document}